\documentclass{appolb}
\usepackage{graphicx}

\usepackage{color}

\begin{document}
\title{Electromagnetic transitions in near-threshold resonances %
\thanks{Presented at the XXXVI Mazurian Lakes Conference on Physics, September 1-7, 2019, Piaski, Poland}%
}

\author{J. Oko{\l}owicz$^{\dag}$, M. P{\l}oszajczak$^{\ddag}$
\address{$^{\dag}$Institute of Nuclear Physics, Polish Academy of Sciences, Radzikowskiego 152, PL-31342 Krak{\'o}w, Poland}\\
\address{$^{\ddag}$Grand Acc\'el\'erateur National d'Ions Lourds (GANIL), CEA/DSM - CNRS/IN2P3,
BP 55027, F-14076 Caen Cedex, France}\\
}
\maketitle

\begin{abstract}
Near-threshold collectivization of continuum shell model eigenstates is investigated in $^{20}$O on the example of B(E$\lambda$) decays of 4$^+$ states in the vicinity of elastic and inelastic neutron threshold. Changes of the electromagnetic transition probabilities as a function of the continuum-coupling strength are explained by the corresponding evolution of the double poles of the scattering matrix.
\end{abstract}
  
\section{Introduction}
Unitarity is the fundamental properties of Quantum Mechanics (QM). Its violation signifies a profound problem in theory. Widely known example is the unitarity crisis in the theory of black hole where no known description of its evolution is consistent with QM \cite{gid2012}.  The mainstream nuclear theory describes atomic nucleus in unitarity violating schemes as the closed quantum system, in glaring conflict with QM. Nucleus is the open quantum system where virtual excitations to continuum states provide an essential mechanism of the effective interaction. Above the lowest particle emission threshold nucleus with a given number of protons and neutrons communicates with other nuclei also by direct particle decays and/or captures. The unitarity crisis of nuclear theory is therefore due to neglecting coupling of discrete and scattering states and may lead to misleading interpretation of several nuclear phenomena. Well known manifestations of nuclear openness are coalescence of eigenfunctions and eigenvalues \cite{EP}, segregation of decay time scales \cite{segr1,segr2}, violation of the orthogonal invariance and channel equivalence \cite{drozdz}, modification of the spectral fluctuations \cite{fyo99}, multichannel effects in reaction cross-section and shell occupancies \cite{hate78}, near-threshold clustering and correlations \cite{oko12}, anti-odd-even staggering of one-nucleon separation energies \cite{luo}, pairing anti-halo effect \cite{benn00}, etc.

Deeper understanding of nuclear properties is provided by the Shell Model (SM) for open quantum systems, such as the Shell Model Embedded in the Continuum (SMEC) \cite{benn00a,oko2003} and the Gamow Shell Model \cite{mic02}. These theoretical approaches allow for spectroscopic studies respecting unitarity in the broad region of masses and excitation energies from drip lines to the region of stable nuclei for states in the vicinity or above the first particle emission threshold. In this work, we will discuss the role of external configuration mixing of SM states through the continuum as a mechanism of the collectivization of near-threshold states \cite{oko12}. This mechanism will be illustrated on the example of electromagnetic transitions from near-threshold SMEC eigenstates in $^{20}$O.

\section{SMEC studies of $^{20}$O}
Numerous examples of of near-threshold resonances have been found in light nuclei \cite{Fick1978,Grancey2016,Wiescher2017,Okolowicz18,Freer2014,Freer2018}. Their frequent appearance must be a general feature, fairly independent of model details \cite{barker1964}. Based on SMEC studies, it has been conjectured \cite{oko12} that the interplay between internal configuration mixing by nuclear interactions and external configuration mixing via decay channels leads to a new kind of near-threshold collectivity. The branch point singularity at the particle emission threshold induces collective mixing of SM states, which results in a single collective eigenstate of the system carrying many characteristics of a nearby decay channel. 
\begin{figure}[htb]
\vskip -3truecm
\centerline{%
\includegraphics[width=9.5cm]{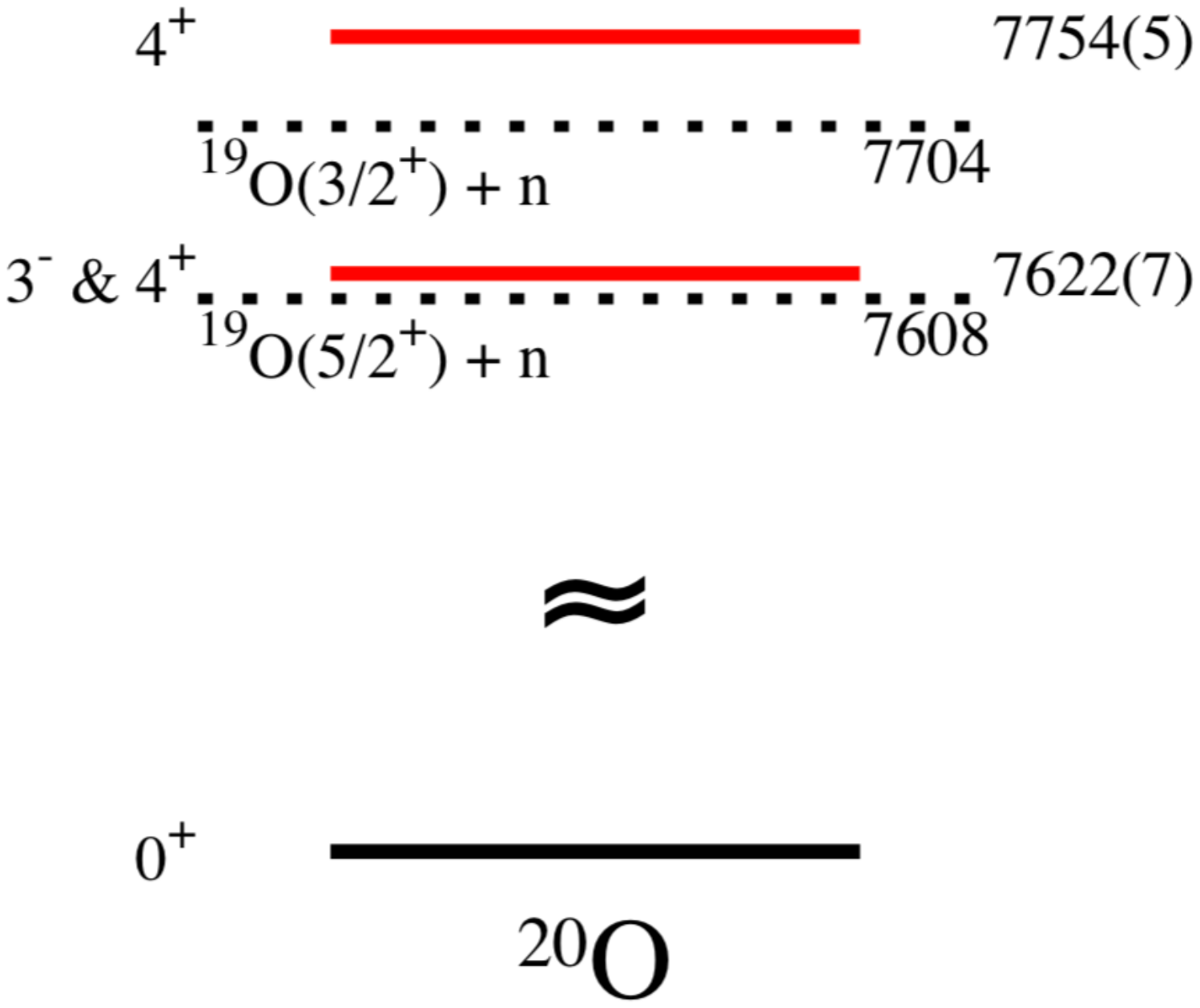}}
\vskip -2.5truecm
\caption{(Color online)  Selected low-lying states and particle-decay thresholds (all in keV) in $^{20}$O \cite{NNDC}.}
\label{Fig:F0}
\end{figure}
Another salient effect is the change of single-particle shell occupancies (spectroscopic factors) and, hence, the modification of NN correlations in near-threshold eigenstates \cite{michel2007}. This effect signifies the change of effective NN interaction in open quantum system eigenstates and is the result of unitarity of the theoretical description. 

Figure \ref{Fig:F0} shows the 4$^+$ states in the proximity of elastic [$^{19}$O(5/2$^+$) + n] and first inelastic [$^{19}$O(3/2$^+$) + n] channels. We shall carry out SMEC calculations to investigate collective coupling of 4$^+$ SM eigenstates to lowest energy neutron decay channels. Detailed description of the SMEC can be found elsewhere \cite{benn00a,oko2003}. In the simplest version of SMEC, Hilbert space is divided into two orthogonal subspaces containing 0 and 1 particle in the scattering continuum, respectively. An open quantum system  description of nucleus includes couplings to the environment of decay channels through the energy-dependent effective Hamiltonian: 
${\cal H}_{\rm eff}(E)=H_{\rm SM}+W(E)$~, where $W$ is the energy-dependent continuum coupling term \cite{benn00a,oko2003} involving couplings between SM eigenstates of $^{20}$O and channel states which are defined by the coupling of one nucleon in the scattering continuum to a SM wave function of $^{19}$O. $E$ in the expression for ${\cal H}_{\rm eff}(E)$ stands for a scattering energy and the energy scale is settled by the lowest one-nucleon emission threshold. The coupling term $W(E)$ induces effective $2N$-, $3N$-, $\cdots$ interactions in the subspace of $A$-particle SM states.

In our study, the SMEC Hamiltonian contains the WBP$-$ interaction \cite{Yuan2017} in the full $psd$ model space. The continuum-coupling interaction is the  Wigner-Bartlett contact force $V_{12}=V_0 \left[ \alpha + \beta P^{\sigma}_{12} \right] \delta\langle\bf{r}_1-\bf{r}_2\rangle$, where $\alpha + \beta = 1$ and $P^{\sigma}_{12}$ is the spin exchange operator. The spin-exchange parameter $\alpha$ has a standard value of $\alpha = 0.73$. The radial single-particle wave functions and the scattering wave functions are generated by the Woods-Saxon (WS) potential, which includes spin-orbit and Coulomb parts. The radius and diffuseness of the WS potential are $R_0=1.27 A^{1/3}$~fm and $a=0.67$~fm, respectively. The strength of the spin-orbit potential is $V_{\rm SO}=5.97$~MeV, and the Coulomb part is calculated for a uniformly charged sphere with radius $R_0$. The depth of the central potential for neutrons is adjusted to obtain the d$_{3/2}$ neutron single-particle state at the measured separation energy of the 4$^+$ state. For protons, the depth of the central potential is chosen to reproduce the measured separation energy of the $p_{1/2}$ orbit. 

\begin{figure}[htb]
\centerline{
\includegraphics[width=6.8cm]{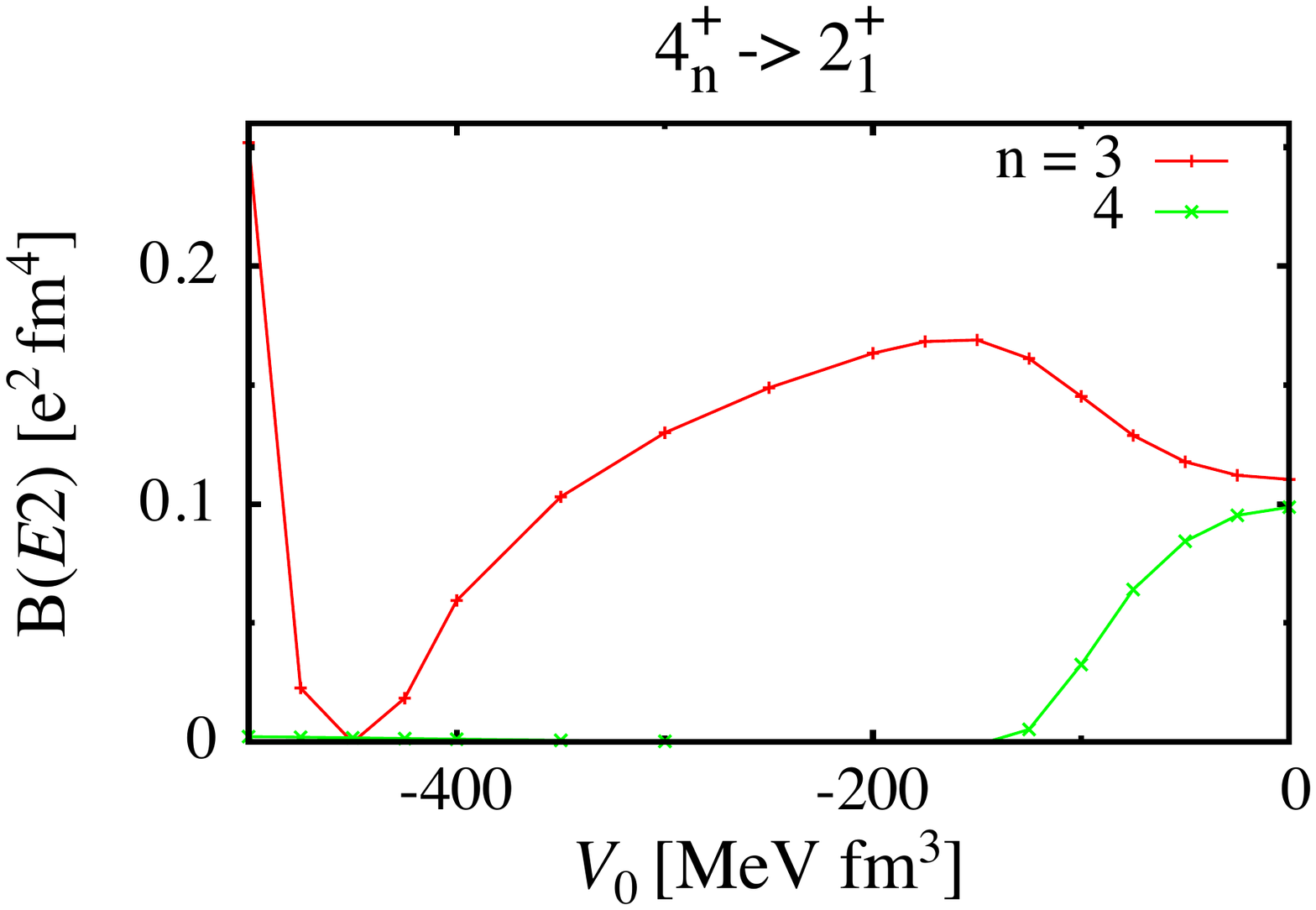}
\includegraphics[width=6.8cm]{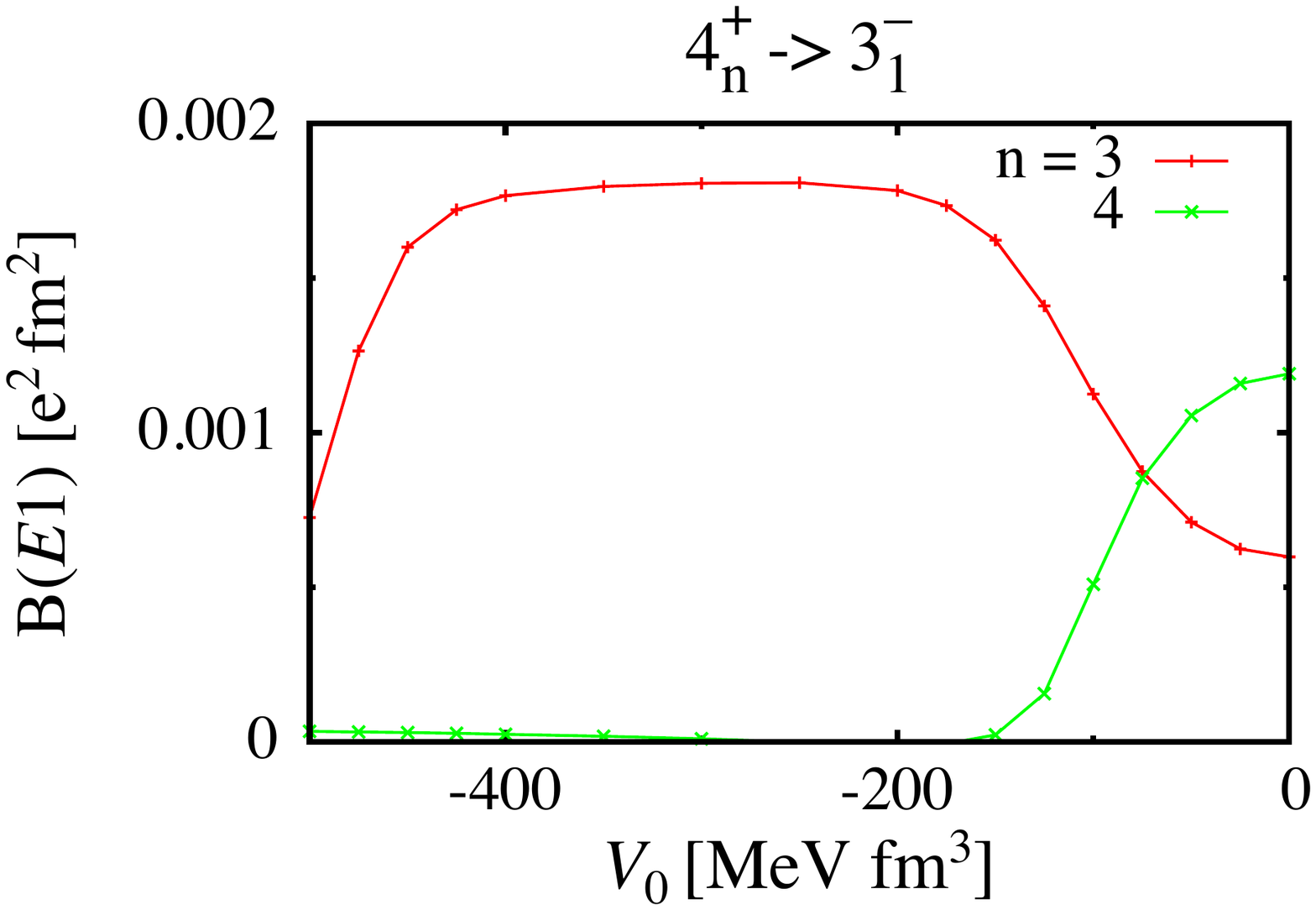}}
\vskip -2.5truecm
\caption{(Color online) $B(E\lambda)$ probabilities in SMEC for the decays of $4^+_n $~ ($n = 3,4$)} resonances in $^{20}$O. Left panel shows $B(E2)$ for $4^+_n \rightarrow 2^+_1$~ ($n = 3,4$) transitions as a function of the continuum-coupling constant $V_0$. Panel on the right hand side exhibits $B(E1)$ for $4^+_n \rightarrow 3^-_1$~ ($n = 3,4$) transitions.
\label{Fig:F1}
\end{figure}
Figure \ref{Fig:F1} displays $B(E2)$ and $B(E1)$ reduced transition probabilities for the decay of near-threshold 4$^+$ resonances to the bound states 2$^+$ and 3$^-$, respectively. SMEC results are plotted as a function of the continuum coupling strength $V_0$. The limit $V_0=0$ corresponds to SM results. The SM states $4^+$ are coupled to the lowest one-neutron decay channels 
\begin{figure}[htb]
\centerline{
\includegraphics[width=6.8cm]{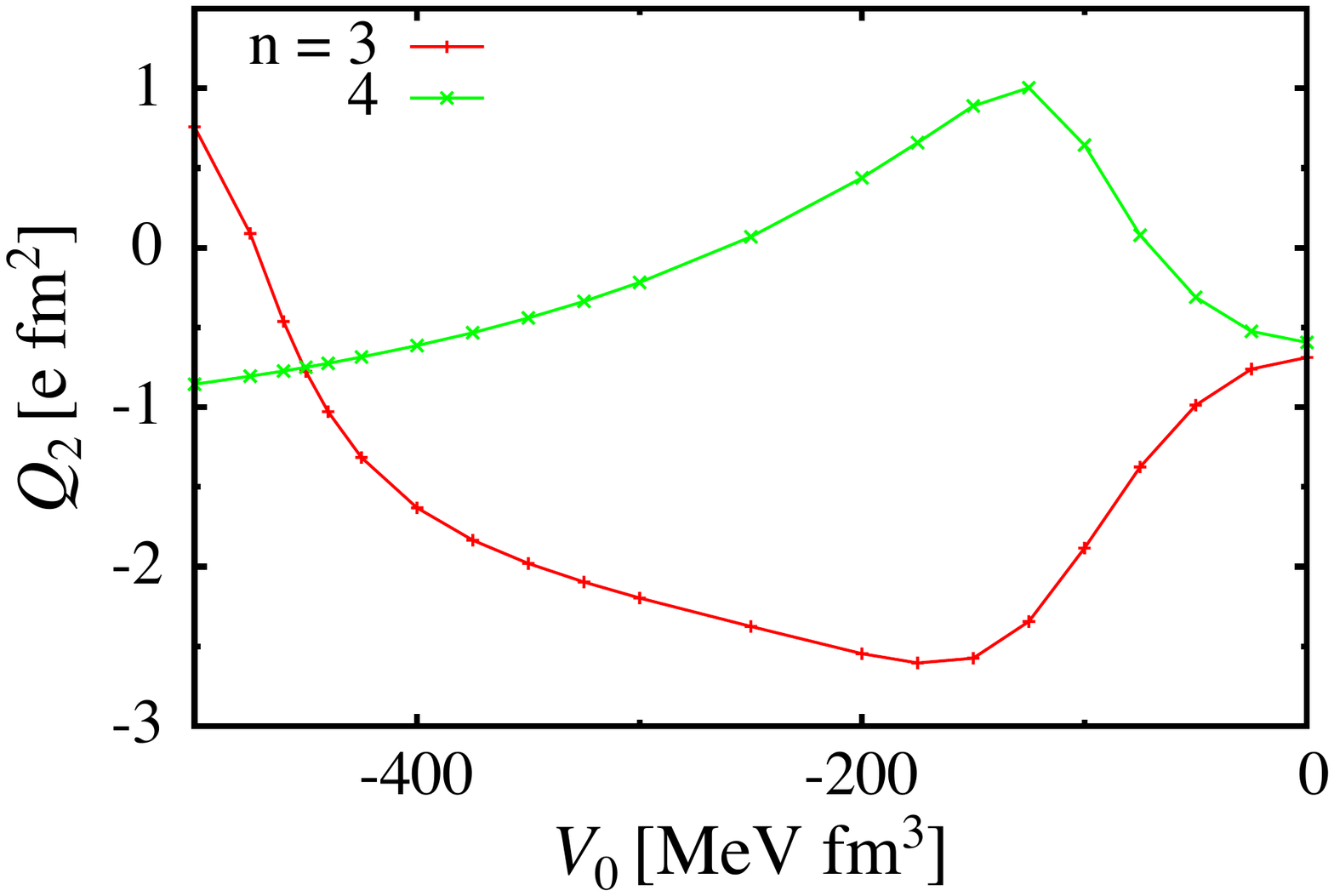}}
\vskip -2.5truecm
\caption{(Color online) Spectroscopic quadrupole moment in SMEC for $4^+_n $~ ($n = 3,4$)} resonances in $^{20}$O are plotted as a function of the continuum-coupling constant $V_0$. 
\label{Fig:F1x}
\end{figure}
$[{^{19}}$O($K^{\pi}_k) \otimes {\rm n}(\ell_j)]^{J^{\pi}}$ with $K^{\pi}_k = 5/2^+_1, 3/2^+_1, 1/2^+_1, 9/2^+_1$, and $7/2^+_1$.
One can notice that for small continuum-coupling strengths, $B(E\lambda)$  for  $4^+_n \rightarrow 2^+_1$~ ($n = 3,4$) and $4^+_n \rightarrow 3^-_1$~ ($n = 3,4$) transitions behave similarly. For $V_0 \leq -150$ MeV fm$^3$, transitions from $4^+_4$ SMEC eigenstate weaken and become close to zero  whereas transitions from $4^+_3$ close to the elastic threshold become stronger. For $-150 \geq V_0 \geq -450$ MeV fm$^3$, $B(E1)$ for $4^+_3$ stays constant whereas $B(E2)$ gradually diminishes and reach zero at $V_0 \simeq -450$ MeV fm$^3$. For very strong continuum couplings ($V_0 \leq -450$ MeV fm$^3$), $B(E2)$ for $4^+_3$ eigenstate grows rapidly whereas $B(E1)$ decreases. 

Figure \ref{Fig:F1x} shows the dependence of spectroscopic quadrupole moment $Q_2$ on the continuum coupling strength for $4^+_3$ and $4^+_4$ resonances. One may notice a significant change of the structure of $4^+$ SMEC eigenstates at $V_0 \simeq -130$  and -440 MeV fm$^3$ associate with the change of the sign of the quadrupole moment. This complicate behavior of reduced transition probabilities is caused by strong mixing of $4^+$ SM states in $4^+$ SMEC eigenstates and can be explained by the proximity of double poles of the scattering matrix, the so-called exceptional points \cite{EP}.

\subsection{Avoided crossings and continuum-coupling correlation energy}

Principal source of the configuration mixing in open quantum systems are the avoided crossings of SMEC eigenstates \cite{oko2003}. Avoided crossings are associated with exceptional points and can be studied by energy trajectories of the exceptional points \cite{EP} of the complex-extended effective Hamiltonian $\tilde{{\cal H}}_{\rm eff}(E)$ in the space of energy and the complex continuum-coupling strength $\tilde{V}_0$. 
\begin{figure}[htb]
\centerline{%
\includegraphics[width=6.8cm]{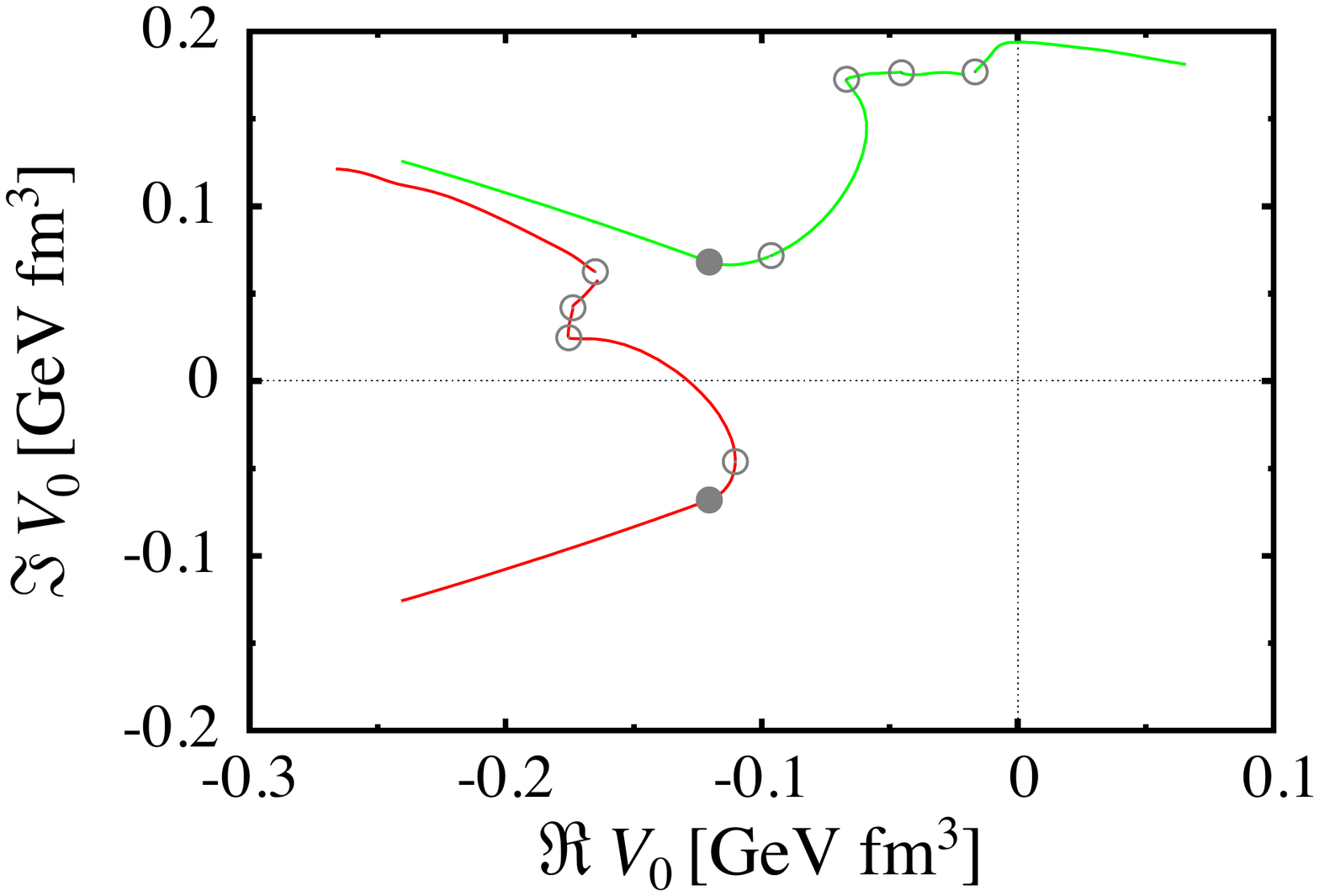}
\includegraphics[width=6.8cm]{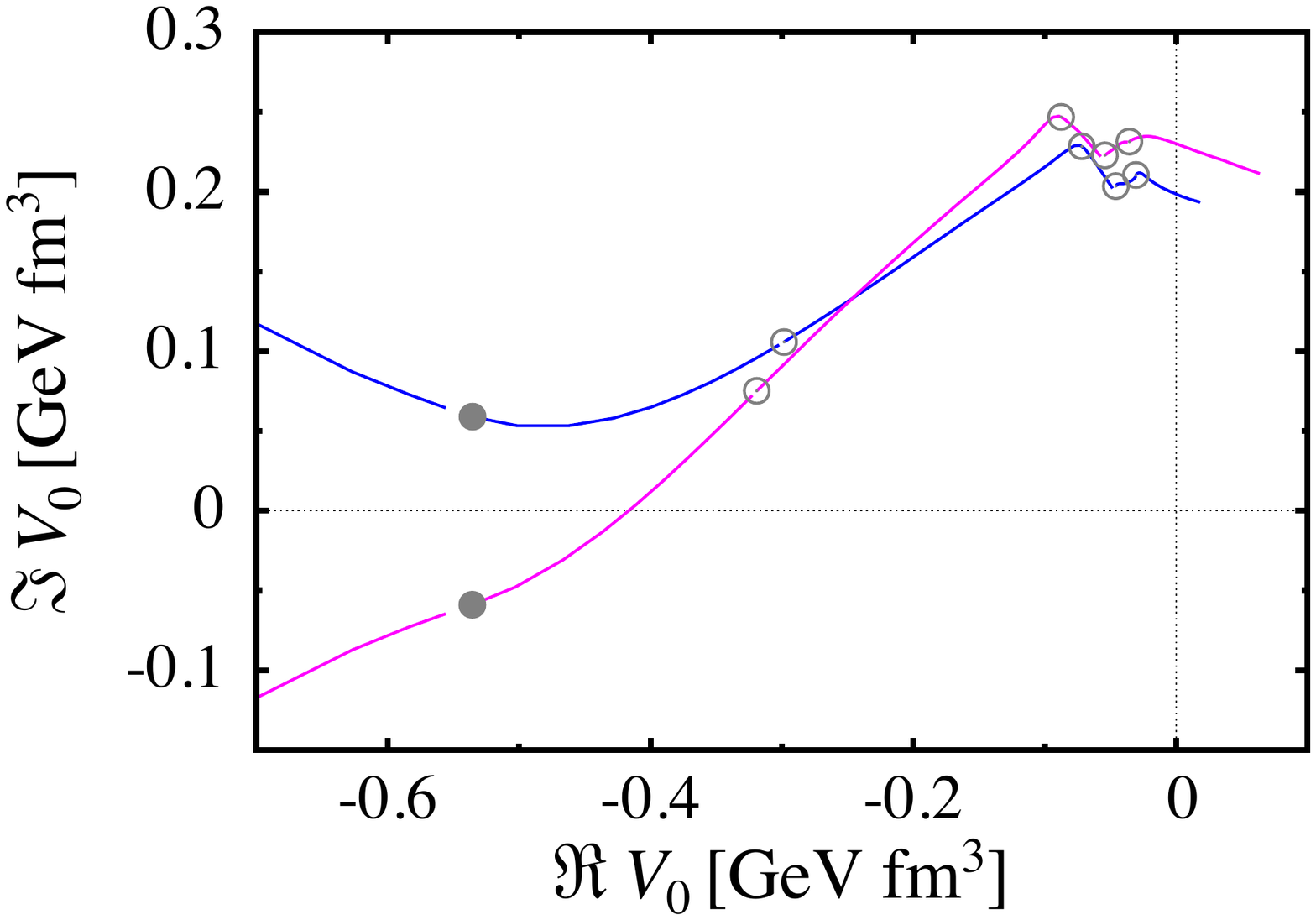}}
\vskip -2.5truecm
\caption{Energy trajectories of the exceptional points for $4^+$ SMEC eigenstates in $^{20}$O are shown as a function of the real and imaginary parts of the continuum coupling strength $\tilde{V}_0$. Negative (positive) imaginary values of $\tilde{V}_0$ correspond to outgoing (ingoing) asymptotics. Different points on these trajectories correspond to different energies $E$. The filled circles denote energies of the elastic threshold. Open circles denote energy thresholds of subsequent inelastic channels. }
\label{Fig:F3}
\end{figure}

Exceptional points are single-root solutions of the two equations:
\begin{equation}
 \frac{\partial^{(\nu)}}{\partial {\cal E}} {\rm det}\left[{\cal H}\left(E;V_0\right)  -{\cal E}I\right] = 0,~~~\nu=0,1.
\label{discr}
\end{equation}
Solutions with both decaying ${\cal I}m(\tilde{V}_0) > 0$ and capturing ${\cal I}m(\tilde{V}_0) < 0$ asymptotics have influence on the configuration mixing of SMEC eigenstates. For a given energy $E$, the maximum number of roots of Eqs. (\ref{discr}) is: $M_{\rm max} = 2n(n - 1)$, where $n$ is the number of states of a given angular momentum $J$ and parity $\pi$. Factor 2 in this expression comes from the symmetry with respect to the transformation $\tilde{V}_0 \rightarrow -\tilde{V}_0$. This symmetry is broken above the lowest particle-emission threshold, {\em i.e.} the analytic continuation of an exceptional point with decaying asymptotics may become a capturing exceptional point, and {\em vice versa}. 

For energies below the threshold of elastic channel, pairs of trajectories shown in left and right panels of Fig. \ref{Fig:F3} are straight lines, reflection symmetric with respect to the axis: ${\cal I}m(\tilde{V}_0) = 0$. In the left panel, one can see trajectories associated with the eigenvalues $4^+_3$ and $4^+_4$. The trajectory which at low excitation energies has decaying asymptotics (see the red curve) crosses the axis ${\cal I}m(\tilde{V}_0) = 0$ at $V_0 = -129$ MeV fm$^3$, $E$ = 428 keV above the elastic reaction channel. Exceptional points along this trajectory have major influence on the values of $B(E\lambda)$ at smaller continuum-coupling constant.  

Trajectories of exceptional points associated with $4^+_3$ and $4^+_2$ SMEC eigenvalues can be seen in the right panel of Fig. \ref{Fig:F3}. 
Here the trajectory associated with the decaying asymptotics (see the curve in magenta) crosses  the axis ${\cal I}m(\tilde{V}_0) = 0$ at 
$V_0 = -417$ MeV fm$^3$, only $E$ = 38 keV above the elastic reaction channel. Exceptional points along this trajectory have major influence on the values of $B(E\lambda)$ for large continuum-coupling strengths ($V_0 \simeq -400$ MeV fm$^3$ and less), leading first to the rapid decrease of both $4^+_3 \rightarrow 2^+_1$ and $4^+_3 \rightarrow 3^-_1$ transition probabilities and then to the  rapid increase of $4^+_3 \rightarrow 2^+_1$ $B(E2)$ probability at $V_0 < -450$ MeV fm$^3$. In this whole range of continuum coupling constants, $B(E2)$ and $B(E1)$ for transitions involving $4^+_4$ eigenstate are negligibly small. One should mention that the double-pole of scattering matrix at the crossing point of magenta trajectory with the axis ${\cal I}m(\tilde{V}_0) = 0$ is outside of the range of relevant values of the continuum couplings. Nevertheless, this exceptional point generates an avoided crossing of $4^+_3$ and $4^+_2$ eigenvalues in the relevant domain of continuum coupling constants 

The continuum-coupling correlation energy:
  \begin{equation}
E_{\rm corr}^{(\alpha)}(E)=\langle \Psi_{\alpha}| W(E) |\Psi_{\alpha}\rangle,
\label{eq22}
\end{equation}
provides complementary information about the configuration mixing and collectivization in a given SMEC eigenstate $\Psi_{\alpha}$.
 Point of the strongest collectivization is determined by an interplay between the Coulomb+centrifugal barrier and the continuum coupling. For higher angular momenta $\ell$ and/or for charged particle decay channels, the extremum of $E_{\rm corr}^{(\alpha)}(E)$ is shifted above the threshold. In our case, the couplings to the decay channels $[{^{19}}$O($K^{\pi}_k) \otimes {\rm n}(\ell_j)]^{4^+}$ is in partial waves $\ell=2$ and 4 for $K^{\pi}_k = 5/2^+_1$ and $K^{\pi}_k = 3/2^+_1$, $\ell=4$ for $K^{\pi}_k = 1/2^+_1$, and $\ell=0$, 2 and 4 for both $K^{\pi}_k = 9/2^+_1$ and $K^{\pi}_k = 7/2^+_1$. 
\begin{figure}[htb]
\centerline{%
\includegraphics[width=6.8cm]{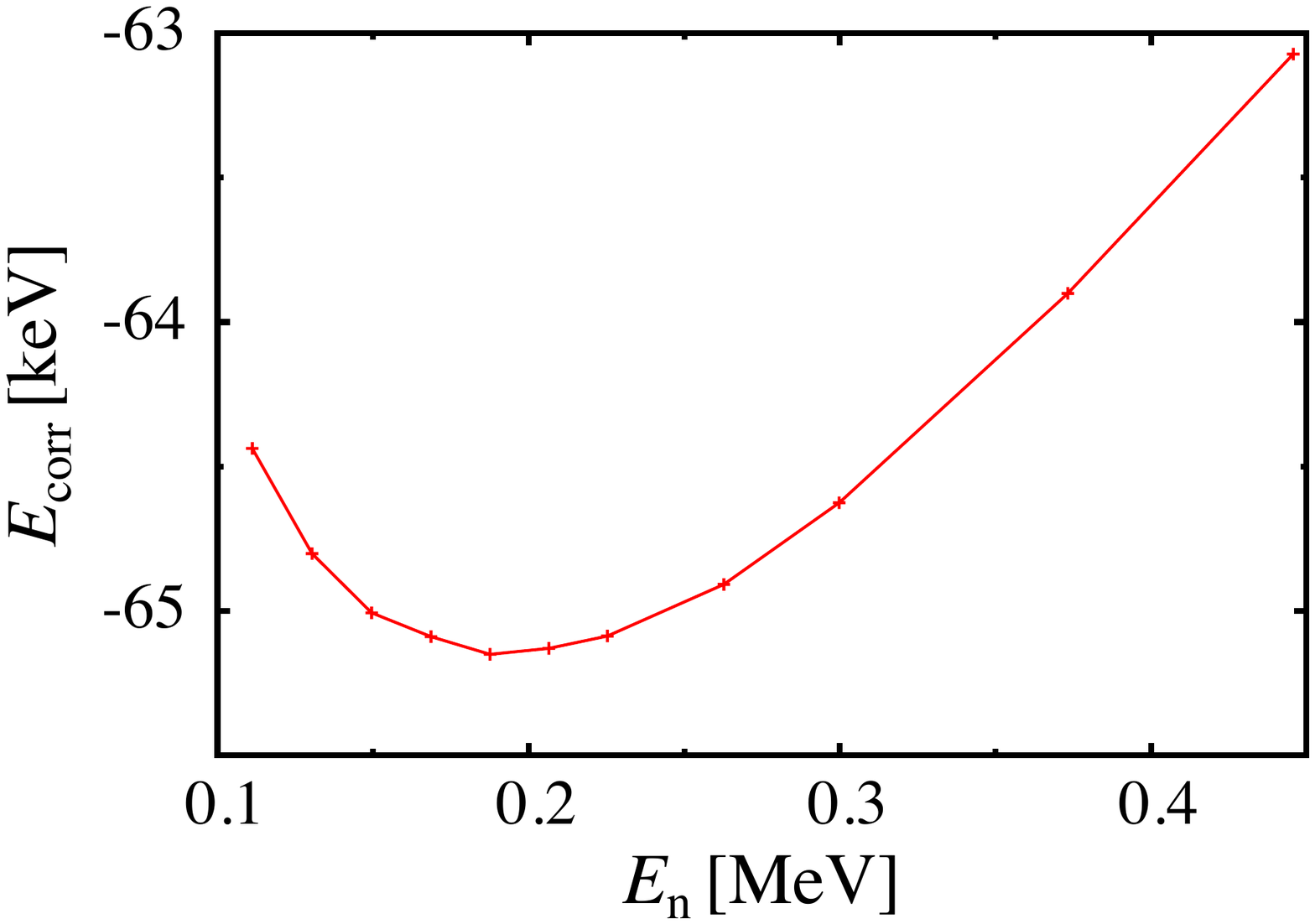}
\includegraphics[width=6.8cm]{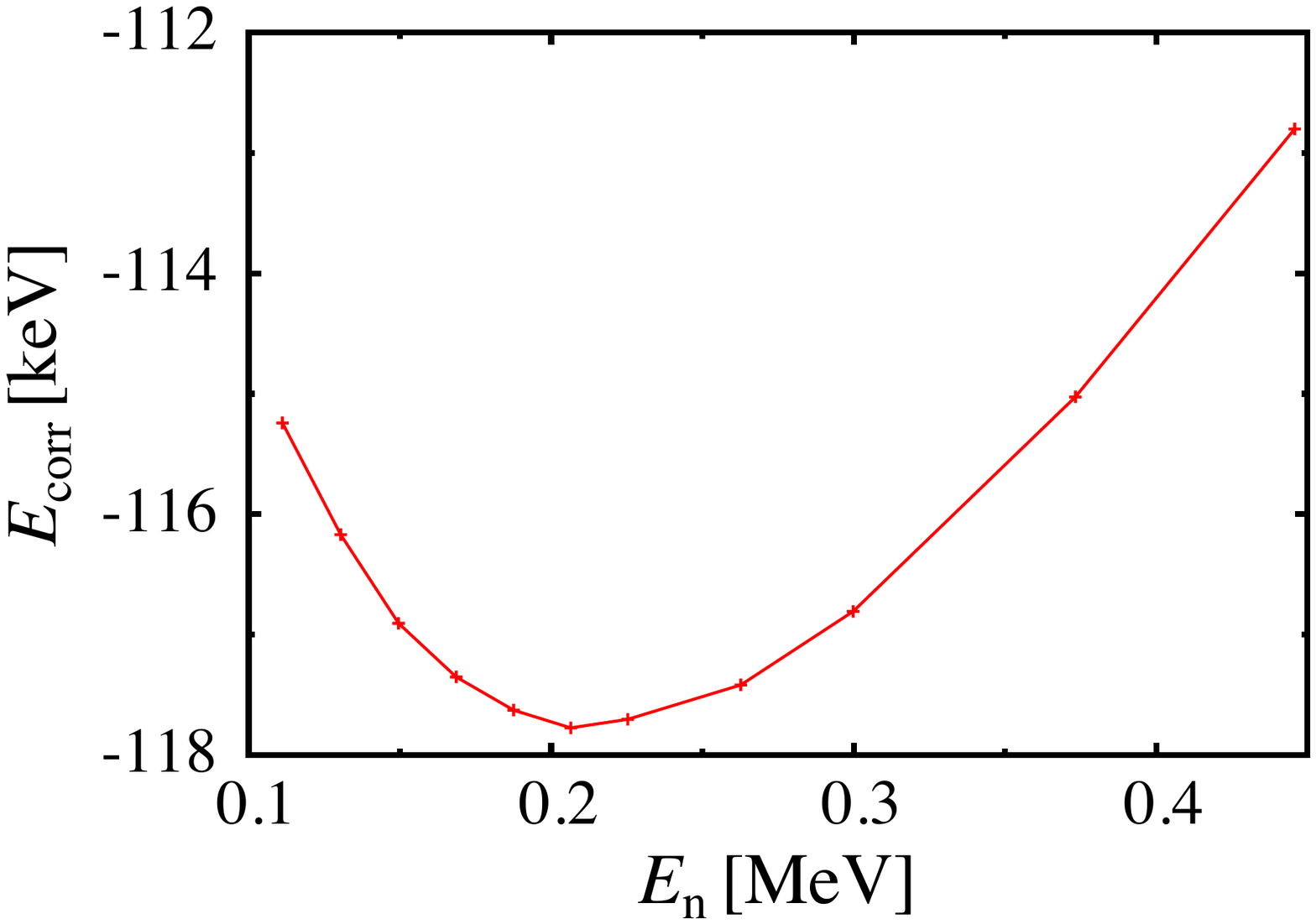}}
\vskip -2.5truecm
\caption{Real part of the continuum-coupling correlation energy for 4$^+_3$ (left panel) and 4$^+_4$ (right panel) SMEC resonances, which are in the vicinity of elastic and lowest-energy inelastic neutron emission thresholds, respectively. The results are shown as a function of the neutron energy $E_n$ in the continuum. Zero energy corresponds to the threshold of elastic channel. Continuum-coupling strength in this calculation is $V_0=-50$ MeV fm$^3$. The Woods-Saxon potential depth for $\ell = 2$ partial wave is adjusted in order to place $d_{3/2}$ single-particle resonance at the energy $E_n$ to ensure proper asymptotics of $4^+$ states.}
\label{Fig:F4}
\end{figure}

The real part of $E_{\rm corr}$ for $V_0 = -50$ MeV~fm$^3$ is plotted as a function of the neutron energy $E_n$ in Fig.~\ref{Fig:F4}. For 4$^+_3$ (left panel) and 4$^+_4$ (right panel) SMEC eigenstates, the corresponding minima appear at $E_n^* \simeq 200$ keV, close to the experimental energy of 4$^+_4$. This resonance is predicted to be strongly collectivized. On the other hand, energy trajectories of exceptional points (see Fig. \ref{Fig:F3}) show the strongest mixing for the pairs of eigenvalues involving 4$^+_3$ because only trajectories ($4^+_3$-$4^+_2$) and  ($4^+_3$-$4^+_4$) cross the axis ${\cal I}m(\tilde{V}_0) = 0$. 
This dichotomy between the information contained in $E_{\rm corr}^{(\alpha)}(E)$ (Fig. \ref{Fig:F4}) and in exceptional point trajectories (Fig. \ref{Fig:F3}) is a result of the complicate multichannel couplings. Indeed, the crossing points with the axis ${\cal I}m(\tilde{V}_0) = 0$ for double-poles ($4^+_3$-$4^+_2$) and  ($4^+_3$-$4^+_4$) appear at significantly different values of $\tilde{V}_0$. With decreasing value of 
$V_0 \equiv {\cal R}e(\tilde{V}_0)$ (see Fig \ref{Fig:F1}), the resonance 4$^+$ close to the optimal collectivization energy $E_n^*$, cease to decay by $\gamma$-emission. 

A similar effect is seen in one-neutron decays. Solution of fixed-point equation \cite{oko2003} at 200 keV above the elastic threshold and $V_0=-50$ MeV fm$^3$ yields $\Gamma(4^+_4)=97.9$ keV and $\Gamma(4^+_3)=44.4$ keV for 4$^+_4$ and 4$^+_3$ resonances, respectively. However, for stronger continuum couplings this tendency is reversed and the 4$^+_3$ resonance becomes significantly broader 
($\Gamma(4^+_3)=2.25$ and 9.3 MeV at $V_0=-200$ and -400 MeV fm$^3$, respectively) whereas 4$^+_4$ shrinks ($\Gamma(4^+_4)=61.3$ and 10.2 keV at $V_0=-200$ and -400 MeV fm$^3$). We see in this example that the near-threshold collectivization of resonance wave functions manifests itself in the scale separation of decay times \cite{segr1,segr2, oko2003}. The resonance 4$^+_3$ which is closest to the elastic reaction threshold becomes superradiant whereas the 4$^+_4$ resonance above the first inelastic threshold is trapped. It is reassuring to notice that these salient effrects of the system openness can be studied both in the $\gamma$- and particle-resonance spectroscopy.

\section{Conclusions}
Near-threshold phenomena are the {\em terra incognita} of nuclear physics. Mixing of SM states via the continuum is at the origin of many new generic phenomena which can be studied in mesosopic open quantum systems, such as atoms, atomic nuclei, atomic clusters, quantum dots, quantum billiards, etc. Uniqueness of these phenomena in atomic nucleus is due to the strong interaction between neutrons and protons which is at the origin of a great variety of particle decays. 

Double-poles of the scattering matrix strongly influence the spectrum and structure of low-energy resonances. In open quantum systems,  location of the double-poles depends strongly on the effective interaction and do not vary in a systematic way from one nucleus to another. From one point of view this poses a tremendous challenge for the microscopic nuclear theory vis-a-vis the microscopic determination of effective nucleon-nucleon interaction. From another point of view, with the systematic data which are sufficiently discriminatory, the continuum coupling constant can be fixed for a given nucleus and the presence of double-pole singularities in the complex-$k$ plane can be deduced from the decay properties of near-threshold resonances. The latter will provide information also about the off-shell behavior of effective NN interaction. 

In the studied case of 4$+$ resonances of $^{20}$O, the energy trajectories of exceptional points and the energy dependence of the continuum-coupling correlation energy give complementary insight into the near-threshold collectivization of open quantum system eigenstates and their decay pattern. We have found a generic phenomenon of the scale separation of resonance decays \cite{oko2003} which in this example leads to the formation of trapped state $4^+_4$ and very broad superradiant $4^+_3$ state in the vicinity of the lowest neutron decay threshold. It is interesting to notice that this effect is seen both in the elctromagnetic and neutron decays which {\em a priori} is not evident. 

Extensive experimental studies are needed to comprehend the rich variety of near-threshold nuclear phenomena and verify the predictions in unitary formulation of the SM.  This is a great future challenge and hope for the nuclear resonance spectroscopy.

\section*{Acknowledgments}
Authors wish to thank Bogdan Fornal and Silvia Leoni for their continuous encouragement and stimulating discussions.This material has been supported by the COPIN and COPIGAL French-Polish scientific exchange programs.

\section*{}

\end{document}